\documentclass[aps,prd,reprint,groupedaddress]{revtex4-2}
\input epsf.sty
\usepackage{graphicx,amsmath,amssymb}
\usepackage{bm}
\usepackage{slashed}
\usepackage{url}
\usepackage{graphicx}

\usepackage{color}
\definecolor{red}{cmyk}{0,1,0.50002,0}
\definecolor{blue}{rgb}{0,0.49995,1}
\definecolor{green}{cmyk}{1,0,0.49998,0}
 
\def\lromn#1{\uppercase\expandafter{\romannumeral#1}}

\begin{document}

%\preprint{magnetization-renp}

%{\blue Comments}


\title{
Implications of neutrino species number
 and summed mass  measurements in cosmological observations
}



\author{N.~Sasao}\email{sasao@okayama-u.ac.jp} 
\author{M.~Yoshimura}\email{yoshim@okayama-u.ac.jp} 
\affiliation{Research Institute for Interdisciplinary Science,
 Okayama University, Tsuhima-naka 3-1, Kita-ku Okayama 700-8530,}

\author{M.~Tanaka}\email{tanaka@phys.sci.osaka-u.ac.jp}%\homepage{https://orcid.org/0000-0001-8190-2863}
\affiliation{Department of Physics, Graduate School of Science, 
             Osaka University,
 Toyonaka, Osaka 560-0043, Japan}

%\author{N. Sasao, %\footnote{sasao@okayama-u.ac.jp} 
%M. Tanaka$^{\dagger}$, and M. Yoshimura} 
%\footnote   
%\footnote{yoshim@okayama-u.ac.jp} 
%\affiliation{Okayama University}
%University \\
%Tsushima-naka 3-1-\affiliation{Research Institute for Interdisciplinary Science,
%Okayama 1 Kita-ku Okayama
%700-8530 Japan  \\
%$^{\dagger}$
%Department of Physics, Graduate School of Science, 
 %            Osaka University\\
 %Toyonaka, Osaka 560-0043, Japan
%}


\date{\today}

%{\color{blue}Changed, to be added or to be modified in blue.}

\begin{abstract}
We confront measurable neutrino  degrees of freedom $N_{\rm eff}$
and summed neutrino mass
 in the early universe  to particle physics at the energy scale  
beyond the standard model (BSM),
in particular including the issue of neutrino mass type distinction.
The Majorana-type of massive neutrino is perfectly acceptable
by Planck observations, while the Dirac-type  neutrino
may survive in a restricted class of models that suppresses
 extra right-handed contribution  to $\Delta N_{\rm eff} = N_{\rm eff} - 3$
at a nearly indistinguishable level from the Majorana case.
There is a chance that
supersymmetry energy scale may be identified in
supersymmetric extension of left-right symmetric model
if improved $N_{\rm eff}$ measurements discover a finite value.
Combined analysis of this quantity with the summed neutrino mass
helps to determine the neutrino mass ordering pattern,
if measurement accuracy of order, $60 \sim 80\,$meV, is achieved,
as in CMB-S4.

%{\color{blue} Statement of IH rejection to be added? }
\end{abstract}

%\keywords{}

\maketitle


%\subsection{}
%\subsubsection{}

%{\bf Introduction} \hspace{0.3cm}

\setcounter{footnote}{0}

\section
{Introduction} %\hspace{0.3cm}

Discovery of non-vanishing neutrino masses by oscillation experiments
requires extension of the electroweak  standard model based on massless
left-handed $\nu_L$  neutrinos.
In general, both left and right handed neutrino fields, $\nu_L^i $ ($ i=1,2,3$ for the three neutrino scheme)
 and $\nu_R^i$ are introduced in this extension. 
Difference of Dirac and Majorana types of neutrino
is in whether projected two-component mass eigenstates have mass degeneracy
(Dirac case), or have two distinct masses 
$m(\nu_R^i)\neq m(\nu_L^i)$ (Majorana case), large and small ones.
The mass degeneracy of Dirac neutrinos implies the law of lepton number conservation,
while the Majorana neutrino violates the lepton number conservation.

The Majorana case provides the interesting seesaw mechanism
explaining why ordinary neutrinos appearing in weak processes are
very light (typically much smaller than $m_{\nu}/m_e = O(10^{-7}$)\, with $m_e$ 
the electron mass) compared to all other fermions, being suppressed
in proportion to $1/m(\nu_R)$ by
heavy $\nu_R$ masses, $ m(\nu_R)$'s \cite{minkowski}.
The lepton number violation has a further consequence  
of interesting  lepto-genesis scenario
which explains the baryon asymmetry of our universe \cite{fy}.
In this scenario the final baryon and the lepton number asymmetries are
comparable in their magnitudes, of order $10^{-10}$, and excludes a possibility of
large chemical potential for leptons.
Experimental determination of  Majorana or Dirac neutrino
is thus one of the most outstanding problems that faces particle physics.

The plausible three-neutrino scheme predicts the neutrino species number 
$N_{\rm eff} =3$ 
 (for a small deviation, see the note \cite{neutrino-spectrum-distortion})
at nucleo-synthesis for three species of Majorana neutrinos.
On the other hand, if
the independent  $\nu_R$ component in the Dirac-type neutrino  
fully contributes to the extra of neutrino species,
the Dirac theory would predict $N_{\rm eff} = 6$ in contradiction
to cosmological observations.
What usually happens is that $\nu_R$ 
decouples earlier from the rest of thermal particles, and after a series
of  subsequent reheating events their number density 
 decreases relative to $\nu_L$ still in thermal equilibrium.
The extent of diluted $\nu_R$ number density is estimated by
the adiabatic relation of entropy conservation,
and it gives an extra contribution to $\Delta N_{\rm eff} = N_{\rm eff} - 3$,
usually time and temperature dependent,
after $\nu_L$ decoupling.

Planck observations of cosmic mircowave background (CMB) 
 provided a stringent value
$N_{\rm eff} = 2.99 \pm 0.17 $ at $1\sigma$ CL \cite{planck-2018}.
This result is perfectly consistent with the Majorana-type neutrino,
but barely consistent with the Dirac-type neutrino.
There have been some discussions of how to understand
this and a less stringent 
 $2 \sigma$ result, $2.99_{-0.33}^{+0.34} $, if the neutrino mass is of
Dirac type since Planck publication \cite{neff-nur1}, \cite{neff-nur2},
\cite{neff-nur3}, \cite{neff-nur4}.
We shall recapitulate the Dirac-type neutrino case
improving calculation of $\nu_R$ decoupling temperature.
Precision $N_{\rm eff}$ measurements in cosmological observations of
CMB-S4 \cite{cmb-s4} is expected to determine this number accurately.

Improved future measurements including null result are expected to
probe species number present in physics beyond the standard model (BSM),
if the neutrino mass of Dirac type.
We find it of particular interest to target 
supersymmetric extension of left-right symmetric model.
In this case it becomes possible to identify SUSY scale if a finite
$\Delta N_{\rm eff}$ is found, or lower energy scale if it is not found.

Our further task in the present paper is a
combined analysis of the neutrino species number and the summed
neutrino mass measurement:
irrespective of Majorana or Dirac neutrinos, the analysis
is shown to have a great impact
on the neutrino mass ordering problem,
the normal hierarchy (NH) or the inverted hierarchy (IH).
We shall be able to provide a new perspective of what
forthcoming observations imply to BSM physics.

The present paper is organized as follows.
In the next section we explain some basic facts about
the neutrino species number $N_{\rm eff}$, and
what cosmological observations of this quantity imply.
The adiabatic entropy conservation is emphasized  and
calculation of the relativistic degrees of freedom
$g_*(T)$ in thermal medium of temperature $T$
 becomes important.
The Majorana-type of massive neutrino is found compatible with
cosmological observations,
while the Dirac-type neutrino requires
new analysis, as done in the literature.
In Section \lromn3 theoretical framework for discussion
of Dirac-type massive neutrino is pointed out.
Calculation of right-handed $\nu_R$ decoupling in this framework is worked out
in detail, since results in the literature lack details of these calculations.
We present detailed explanation of $\nu_R$ production, adding
processes not considered in the literature.
It is shown that four-Fermi type approximation of $\nu_R$ pair production rate
is sufficient to determine the thermalization condition 
and the decoupling temperature.
In Section \lromn4 we discuss how the derived decoupling
temperature is related to diluted $\Delta N_{\rm eff}$
and the necessary dilution  may be realized in unification schemes.
We may relate  SUSY energy scale to $\Delta N_{\rm eff}$
and shall be able to discuss how supersymmetric left-right symmetric schemes
are constrained.
In Section \lromn5 another important quantity of summed neutrino masses
in future cosmological observations may be combined to neutrino
oscillation data, and resulting plot in $(\Delta N_{\rm eff}, \sum_i m_i)$
plane can be used to determine another neutrino property,
normal and inverted mass hierarchical ordering.

Throughout this paper we use the  unit of $\hbar = c = k_B = 1$.

\section
{Preliminary}

\subsection
{$N_{\rm eff}$ at nucleo-synthesis and at epochs after recombination}
%\hspace{0.3cm}

The effective massless degrees of freedom  $N_{\rm eff}$
determines the cosmic energy density, hence controls the speed of cosmic expansion.
This quantity at nucleo-synthesis ($\sim 100$ seconds since the big-bang) is measurable 
by comparing measured light element abundances
with theoretical calculation, 
while the same quantity at  later epochs after
recombination ($\sim 4 \times 10^5$ years after the bang)
is measured  by Planck and CMB-S4 observations.
The quantity $N_{\rm eff}$ at these two epochs is sensitive function
of $^4$He abundance $Y_p$.
The allowed  region in the $(Y_p, N_{\rm eff})$ plane from observations
 is usually presented by contour maps.
The important fact is that correlations given by the  
derivative signs of  $dN_{\rm eff}/dY_p$ 
are opposite at two epochs, positive at nucleo-synthesis
and negative at recombination.
This makes it possible to determine both of $N_{\rm eff}$ and $Y_p$ at high precision.

The primordial $^4$He abundance $Y_p$ is around 0.25.
Reference \cite{dolgov:report} on nucleo-synthesis cites a conservative
bound $N_{\rm eff} < 4$, but also mentions a more restrictive bound 
$N_{\rm eff} < 3.2$.
The correlation with $Y_p$ at nucleo-synthesis gives 
a straight line in the $(N_{\rm eff}, Y_p)$ plane, as reviewed in 
\cite{cmb-s4}, ranging in $N_{\rm eff} = 2.2 \sim 3.4$.
The final status of Planck 2018 + BAO  observations gives an impressive
upper bound $N_{\rm eff} < 3.33 $ at 95\%  CL \cite{planck-2018}.

Not only neutrinos, but also other stable light %(masses less than 0.1 eV)
 relics left behind from the early thermal history give additional contribution,
 extra $\Delta N_{\rm eff} = 0.03 \sim 0.05$ per a single relic.
Examples are axion, dark photon and sterile neutrino.
Contribution from a spin-less stable light relic is $\Delta N_{\rm eff} \sim 0.027 $.
This quantity can also be negative if light relics are unstable and decay during two epochs.
We shall assume that there is no such relic, or even if there is,  their
cumulative  contribution is
restricted at most by $| \Delta N_{\rm eff}| < 0.028$
(CMB-S4 target value), and concentrate
on diluted Dirac $\nu_R$ which may behave in a similar way to a light relic.

\subsection
{Adiabatic dilution factor and $\Delta N_{\rm eff}$ }
%\hspace{0.3cm}

Assume  that $\nu_R$ in the Dirac theory is thermally abundant
at early cosmological epochs,
and calculate the amount of dilution. 
After  $\nu_R$ decoupling the universe goes through many
annihilation events of thermal anti-particles, thereby reheating
the universe. 
Electroweak interaction is far stronger
than interactions in BSM, hence
left-handed $\nu_L$ is reheated, but $\nu_R $ is not.
This gives rise to difference of $\nu_L$ and $\nu_R$
effective temperatures. Note that
even after $\nu_R$'s decouple and become non-thermal, 
one can assign their effective temperature.
This is because the  entropy conservation of $a^3 s(T)$ (with $a(t)$ the
time-dependent cosmic scale factor) holds at nearly instantaneous
reheating processes.
 The entropy density $s(T)$ is proportional to
the massless degrees of freedom $g_*(T)$ in radiation-dominated epoch.
Hence one can readily calculate the temperature ratio before and after
rehearing events by counting respective species number contributing 
to $g_*$.

Assuming that the standard electroweak phase transition took place in the Big Bang Universe,
it is straightforward to evaluate $g_*$ in the era up to the electroweak phase transition in the standard model.

Fermions are counted by the weight $7/8$ for one spin state
and bosons by the weight $1$.
Counting  $g_*(T)$ gives the  dilution factor in terms of number density ratio, 
$n(\nu_R)/n(\nu_L) = 43/427= 0.101$,
since the electroweak phase transition.

We summarize $g_*$  factors relevant to $\nu_R$ dilution in the following table, Table(\ref{table 1}) in which $g_*^{\rm i}$ indicates the species number at cosmological events i.
%hspace*{0.5cm}
\begin{table}
\begin{tabular}{|c|c|c|c|c|}
\hline
i & $\nu_L$  & EW & $\nu_R$ &  S4\\ 
\hline 
$g_*^{\rm i}$ & 10.75 & 106.75 & 124.75 & 360\\
\hline
$\Delta N_{\rm eff}^{\rm i} $ & 3 & 0.14 & 0.11 & 0.028 \\
\hline
\end{tabular}
\caption{
Species number $g_*^{\rm i}$ with i indicating  cosmological events at which $\nu_R$ decoupling occur simultaneously (in an approximate sense).
i $= \nu_L$ being left-handed neutrino decoupling, EW  standard model phase transition, $\nu_R$  the phase transition of the doublet left-right symmetric model, while $g_*^{\rm S4}$ is the expected species number that can be probed by CMB-S4. The formula $3 (g_*^{\nu_L}/g_*^{\rm i} )^{4/3} $ is used for $\Delta N_{\rm eff}^{\rm i}  $.
} 
\label {table 1}
\end{table} 
%\vspace{0.5cm}
It is convenient in the  rest of analysis to relate the extra 
$\Delta N_{\rm eff}$ to the extra  relativistic degrees of freedom
$\Delta g_*$,
\begin{equation}
\left(  \frac{ g_*^{\rm EW} + \Delta g_*}{ g_*^{\nu_L}}\right)^{- 4/3} 
= \frac{\Delta N_{\rm eff}}{3} 
\,,
\label {neff-g* relation}
\end{equation}
where $g_*^{\rm EW} = 427/4$ is the species number of electroweak theory, and
$g_*^{\nu_L}= 43/4 $. 
Note that $\Delta g_*$ can be negative, implying that
$\nu_R$ decoupling may occur below the electroweak temperature 
$T^{\rm EW} \sim 250\,$GeV.

Alternatively, one can relate $\Delta N_{\rm eff}$ upper bound
or observed value to required $\Delta g_* $ value, using
the formula (\ref{neff-g* relation}).

\section
{Fate of right-handed neutrino of Dirac type}

\subsection
{Theory of Dirac-type massive neutrino}

Let us first recall what happens in the standard electroweak theory
based on the gauge group SU(2)$_L \times $U(1), when one introduces
the right-handed neutrino $\nu_R$.
It is known  that introduction of SU(2)$_L \times $U(1) singlet $\nu_R$
opens the possibility of Higgs boson ($h$) coupling 
proportional to doublet bi-linear fermion $(\bar{\nu_L}, \bar{l}_L) \nu_R$
which, after the spontaneous electroweak gauge symmetry breaking,
gives the Dirac-type neutrino mass.
Inevitable process $\nu_L h \rightarrow \nu_R$ from thermal $\nu_L, h$
produces $\nu_R$, but with
 a negligible amount of  $\nu_R$ ($\Delta N_{\rm eff} \ll 0.05$) if
neutrino masses are  less than of order 100 eV \cite{dolgov:report}.
There seems nothing wrong with this scheme, but
this minimum extension does not give any clue to
many outstanding problems of particle physics such as
generation of the baryon asymmetry.
We need a new theoretical framework of massive Dirac-type neutrino.

A natural scheme is grand unified gauge theories (GUT), but
it is sufficient to first think of an intermediate
step of subgroup unification towards GUT.
We find it most natural to analyze $\nu_R$ dilution
in the left-right symmetric extension of the standard electroweak theory,
SU(2)$_R \times$SU(2)$_L \times $U(1) gauge theory \cite{minimum_su(2)xsu(2)xU(1)},
as also made in \cite{neff-nur3}.
The Higgs system consists of irreducible representations, 
$(2,2)\,, (2,1)\,, (1,2)$ of SU(2)$_R \times$SU(2)$_L$ group. 
Regarding all species of particle as massless, there are 18 extra $g_*$ in addition to
the standard one $427/4$ and three right-handed neutrinos $21/4$.

Some details of this doublet left-right symmetric model (DLRSM) are given in Appendix.

\subsection
{Decoupling of thermalized right-handed neutrinos
in LR symmetric model}

The Boltzmann equation for the number density, obtained after
integrating the distribution function $f(\vec{p})$ in the phase space
(space volume times the momentum-space volume $d^3 p/(2\pi)^3$ in
spatially homogeneous universe), 
 describes time evolution in the expanding universe.
Consider a number density of $f$ species $n_f$. The equations is
\begin{eqnarray}
&&
\dot{n}_f + 3 H(T) n_f = \sum_i \Gamma_i n_{i \neq f}- \Gamma_f n_f 
\,,
\label {boltzmann eq}
\end{eqnarray}
with $H(T)$ the Hubble rate $\sqrt{\frac{8\pi G_N}{3} \rho_r} $
($\rho_r$ the energy density of effectively massless particles,
$\pi^2 g_* T^4/30$).
When the Hubble rate $H(T)$ compared with right-hand side (RHS) is small, 
thermal equilibrium is
realized with vanishing RHS of (\ref{boltzmann eq})
 (meaning the positive production and the negative
destruction balance).
This thermal equilibrium is ended at decoupling temperature $T_d^f$
that is  determined by equating the rate to
the Hubble rate, $H (T_d^f) = \Gamma_R $. 
%The Hubble rate  is given as $H(T) \approx g_*(T) T^2/m_{\rm pl}\,, m_{\rm pl}= 1.22 \times 10^{19}\,$GeV (the Planck mass).
One may use for $\nu_R$ annihilation rate $\Gamma_R$ 
\begin{eqnarray}
&&
\hspace*{-0.5cm}
\Gamma_R(T) = \frac{ 16}{n_R(T) } \int \frac{d^3 p_1 d^3 p_2}{ 4 p_1 p_2 (2\pi)^6}
%\nonumber \\ && \hspace*{0.3cm} \times 
\frac{{\cal S} }{(e^{p_1/T} + 1) (e^{p_2/T} + 1) }
\,,
\label {nu-r annihilation rate}
%\nonumber 
\\ &&
\hspace*{-0.3cm}
{\cal S} =
 \int \frac{d^3 q_1 d^3q_2 }{ 4 q_1 q_2 (2\pi)^2 }
\delta^{(4)} ( p_1 + p_2 - q_1 -q_2)
\, {\cal R}(s,t) 
\,.
\label {nu-r annihilation rate 2}
\end{eqnarray}
Momenta of initial and final particles, assumed massless,
 are denoted $p_i\,, q_i \,, i=1,2$, respectively.
Lorentz-invariant squared amplitude ${\cal R}(s,t)$ 
summed over spin states is written in terms of invariant variables,
 $s=(p_1+ p_2)^2 \,, t = (p_1 - q_1)^2$.
They are calculated below.
$n_R(T) = 3\zeta(3)(2s_p+1) T^3 / 4\pi^2$ is the number density of initial
thermal $\nu_R$ ($s_p=1/2$ is its spin).

One may separately confirm that thermalization condition is satisfied
by considering inverse processes.
Since we treat all initial and final fermions as massless (actually
much lighter than gauge bosons $W_R, Z_R$)
inverse $\nu_R$ production process has equal rate to the annihilation rate.
Thus, time integrated quantity $ n_R \Gamma_R $ gives thermally summed
$\nu_R$ number density.

\begin{figure*}[htbp]
 \begin{center}
 \epsfxsize=0.8\textwidth
 \centerline{\epsfbox{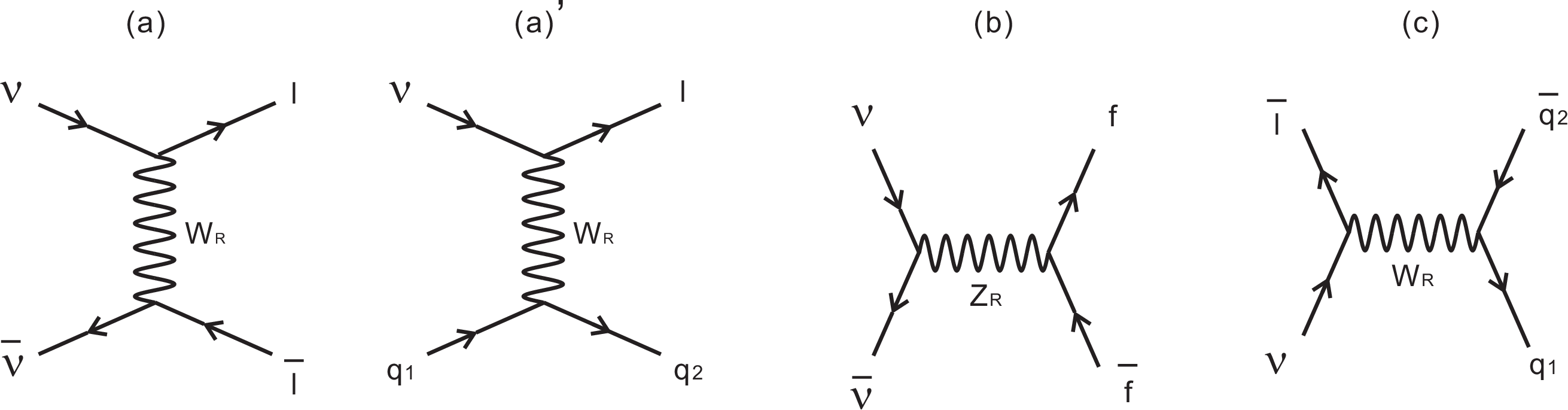}} \hspace*{\fill}\vspace*{1cm}
   \caption{
Typical Feynman diagrams: (a) $ \nu_R \bar{\nu}_R  \rightarrow l_R \bar{l}_{R'}$
(t-channel $W_R$ exchange), (a') 
$\nu_R q^1_R  \rightarrow l_R q^2_R \,, (q_1, q_2) =
(d,u), (s,c), (b,t )$ (t-channel $W_R$ exchange),
(b) $ \nu_R \bar{\nu_R}  \rightarrow f_R \bar{f_R}
\,, f = q, l$ (s-channel $Z_R$ exchange),  (c)
additional s-channel $W_R$ exchange contribution,  
$\nu_R \bar{l_R}  \rightarrow q^1_R \bar{q}^2_R \,, (q_1,q_2) = (u,d), (c,s), (t,b)\,,
l = e, \mu, \tau$.
}
   \label {feynman 1}
 \end{center} 
\end{figure*}

Relevant annihilation processes of right-handed neutrino  $\nu_R$
(conjugate processes of $\bar{\nu}_R$ production to be included as well)
in LR symmetric models occur via two-body colliding processes of many kinds.
Typical Feynman diagrams are depicted in Fig(\ref{feynman 1}).
In order to calculate rate, we classify individual contributions as
\begin{eqnarray}
&&
(a); \; 
(al);\;
\nu_R \overline{\nu_R'} \rightarrow l_R \overline{l_R'} \,, \; 
(aq):\;
\nu_R \overline{u_R}  \rightarrow l_R \overline{d_R} \,,
\nonumber \\ &&
\hspace*{0.5cm}
( t{\rm -channel} \; W{\rm -exchange})
\,, 
\label {diagram 1}
\\ &&
(a)'; \; 
\nu_R d_R  \rightarrow l_R u_R
\,, \;
( t{\rm -channel } \; W{\rm -exchange}),
\\ &&
(b); \; 
\nu_R \overline{\nu_R}
\rightarrow f_R \overline{f_R} 
\,, %\,, f = q, l
\; ( s{\rm -channel } \; Z{\rm - exchange})
\,,
\\ &&
(b)'; \; 
\nu_R \overline{\nu_R}
 \rightarrow f_L \overline{f_L}
\,, %\,, f = q, l
\; ( s{\rm -channel } \; Z{\rm- exchange})
\,,
\\ &&
(c); \;
\nu_R \overline{l_R} 
 \rightarrow u_R \overline{d_R} 
\,, %\,, f = q, l
\; ( s{\rm -channel } \; W{\rm- exchange})
\,,
\label {diagram 5}
\end{eqnarray}
 with $f= q\,, l$ being quarks and charged lepton.
We note that process involving light higgs pair  is absent, because
a neutral scalar field has no vector current that may couple to $Z_R$.
Feynman rules for amplitude and rate calculation can readily be
extracted using formulas in Appendix.

A part of these contributions to 
annihilation rates have been calculated in the literature.
Their substantial part is missing in the literature, and we shall cover
and add all relevant contributions.
The decoupling temperature $T_d$ at which two rates are equal is
related to right-handed gauge coupling masses and gauge coupling $g_R$.
Remarkably, we shall be able to provide analytic results for
important quantities we need.

SU(2)$_R \times$SU(2)$_L \times $U(1) gauge theory gives
contributions from $(a) \sim c)$ listed in (\ref{diagram 1})
$\sim $ (\ref{diagram 5}), some of them
depicted  in Fig(\ref{feynman 1}).
Invariant squared amplitudes ${\cal R}(s,t)$ after spin summation consist of
coupling factors and dynamical parts given in terms of $s,t$ variables.
Listed contributions have different $s,t$ dependence.
It is sufficient to calculate ${\cal R}(s,t)$
in the temperature range $T \ll $ gauge boson masses, hence
four-Fermi approximation is excellent, with the common strength factor $G_R$,
\begin{eqnarray}
&&
\frac{g_R^4} {M_Z^4 \cos^2 \theta_R} = \frac{g_R^4} {M_W^4} \equiv
32\,G_R^2
\,.
\end{eqnarray}
We used notations  $M_W\,, M_Z$ for $W_R, Z_R$ masses,
to distinguish from ordinary electroweak gauge bosons $m_W, m_Z$, 
and $g_R\,, \theta_R$ are gauge coupling constant and mixing angle
in the SU(2)$_R$ sector.
Flavor dependent coupling factors 
 are further multiplied to squared Fermi constant;
\begin{eqnarray}
&&
C_{2t}^{al}  = \frac{3}{4} 
\,, \hspace{0.3cm}
C_{2t}^{aq}  = \frac{9}{4} 
\,, \hspace{0.3cm}
C_{2t}^{a'}  = \frac{9}{4} 
\,,\\ &&
C_{2s}^{ b} = 
\frac{1}{4}
\left( 9 (\frac{1}{2} - \frac{2}{3} \sin^2 \theta_R)^2 
+ 9 (-\frac{1}{2} + \frac{1}{3} \sin^2 \theta_R)^2 
\right. \nonumber
\\ && \hspace{10mm}
\left.
+ 3 (-\frac{1}{2} + \sin^2 \theta_R)^2 
\right)
\,,
\\ &&
C_{2s}^{b'} = \frac{1}{4}
\left( 18 ( - \frac{\sin^2 \theta_R}{6})^2 + 6 (  \frac{\sin^2 \theta_R}{2})^2
\right)
\,,
\\ &&
C_{2s}^{ c} = \frac{9}{4}  
\,, \hspace{0.3cm}
C_{st} =    \frac{1}{4} ( -\frac{1}{2} + \sin^2 \theta_R)
\,.
\end{eqnarray}
Contributions $C^{\alpha}_{\varphi}$ in squared amplitudes
arise from $\alpha-$type of diagrams 
in Fig(1) and those of $\varphi-$type of exchanged gauge bosons.
For instance, $C^{aq}_{2t} $ is from squared $t-$channel exchange
of Fig(1a) in which leptons $l$ are replaced by relevant quarks $q$.

The last coupling $C_{st}$ arises from interference contribution of 
t-channel $W_R$-exchange (a) diagram and s-channel $Z_R$-exchange (b) diagram
for two-body process of
$\nu_R \bar{\nu}_R \rightarrow l_R \bar{l}_R  \,, l=e, \mu,\tau $.
As explained in Appendix,
a value $\sin^2 \theta_R= 0.299$ should be used.
Using these couplings, squared invariant amplitudes are given by
\begin{eqnarray}
&&
%\hspace*{-0.3cm}
{\cal R} (s,t)= 32 G_R^2 \left(
C_{2t}^{a'}  s^2 + (C_{2t}^{al}+C_{2t}^{aq}) (s+t)^2 
\right.
\nonumber \\ &&
\hspace*{-0.5cm}
\left.
+ C_{2s}^{ b}(s+t)^2 
+ C_{2s}^{b'}t^2  + C_{2s}^{ c}(s+ t)^2
- 2 C_{st} (s+t)^2
\right)
\,.
\end{eqnarray}

The last integral over final states in (\ref{nu-r annihilation rate 2}) is to
be calculated in general coordinate frames.
If one  replaces this  by the total cross section in the
center-of-mass frame,
the important part of asymmetric collisions
between initial fermion pairs is lost.

The angular integration over final state variables is done
in Lorentz-invariant manner,
which gives trivial angular integrations,
$\int (s+t)^2 = \int t^2 = \frac{1}{3} s^2\,, \int s^2 = s^2$.
Thus, the annihilation cross section is
\begin{eqnarray}
&&
\sigma v = 4 \frac{1}{2p_1p_2} \frac{1}{16\pi\,s } \, \int_{-s}^0 dt\, {\cal R}(s,t)
\nonumber \\ &&
\hspace*{0.5cm}
= \frac{4 G_R^2}{ \pi} \frac{s^2}{p_1 p_2} \Big( 
C_{2t}^{a'} + \frac{1}{3} (C_{2t}^{al}+C_{2t}^{aq}
\nonumber \\
&& \hspace{20mm}
 + C_{2s}^{ b}+ C_{2s}^{b'} + C_{2s}^{ c}
 - 2 C_{st}) \Big)
\,.
\end{eqnarray}
Using $s = 2p_1 p_2 ( 1- \cos \theta_{12})$,
the angular  integral in the initial state
gives $ 4 \pi^2 \int_0^{4 p_1 p_2} ds/p_1 p_2$.
Finally, 
integration over initial thermal fermions leads to
\begin{eqnarray}
&&
\Gamma_R = \frac{ 49 \pi^5}{3 \zeta(3)\cdot 675} \, G_R^2 T^5 
\, I(\sin^2 \theta_R)
\nonumber \\ && 
= 
26.2 \, G_R^2 T^5
\,, \hspace{0.3cm}
I(x) = \frac{1}{48}( 217 - 56 x + 40 x^2)
\,,
\label {nur annihilation rate}
\end{eqnarray}
using the value of
$x= \sin^2 \theta_R = 0.299$, given in Appendix.

We presented results using exact Fermi-Dirac (FD) distribution function for
fundamental fermions,  quarks and leptons.
In Appendix, we also derive results taking the approximate Maxwell-Boltzmann (MB)
distribution, which gives  analytic results 
that differ by
\begin{eqnarray}
&&
\left( \frac{\int_0^{\infty} dx x^3 e^{-x} }{\int_0^{\infty} dx x^3/(e^x+1) } \right)^2 =
\frac{518400 }{49 \pi^{8} } = 1.11499
\,,
\end{eqnarray}
The result of annihilation rate divided by $n_f$ is
$0.9948 $ times FD value, two values being
surprisingly close to each other within 1 \% difference.

Some comments on thermal distribution function may be in order.
After copious production right-handed neutrinos scatter with thermal
particles (via $Z_R$ exchange) and their energies are redistributed.
Thus, $\nu_R$ thermal distribution function is realized with
zero chemical potential.
Thermal gauge bosons $W_R$ can produce $\nu_R$, but
they decouple much earlier, resulting in no further thermal $\nu_R$
production.

We have also calculated $\nu_R$ production rates in a
much more simplified approximation
of using thermally averaged values in the  invariant total cross section $\sigma v$.
Using the averaged $\bar{t} = - s/2\,, \bar{t^2} = s^2/3$,
one has the thermally averaged $\sigma v = \overline{{\cal S}}/4p_1 p_2  $ given by
\begin{eqnarray}
&&
\sigma v =
\frac{4 G_R^2}{\pi} s  I(\sin^2 \theta_R)
\,.
\end{eqnarray}
This gives $\Gamma_R \approx 6 \zeta(3)
 G_R^2 \bar{s} I /\pi^3$.
The thermal average of $s$ is $ \bar{s} = 2 \bar{p}^2 = 1.814\,T^2$, hence
one derives  $\Gamma_R \sim 1.79 G_R^2 T^5$, 
which grossly differs from
the more precise result (\ref{nur annihilation rate}).
This estimate confirms the importance of asymmetric collision in rate
calculations.
Note that one is not allowed to take
the center of mass frame in the comoving
frame of thermal universe, since head-on collisions do not necessarily occur
and collisions with angles do occur usually. 
Thus, one has to perform thermal average over angular configurations.
The invariant variables $s= 2E_1 E_2 (1-\cos \theta_{12})\,, 
t = - 2E_1 E_2 (1-\cos \theta_{13})$ 
of a massless $(1,2)$ pair collision $(1,2)\rightarrow (3,4)$
 must be averaged using thermal distribution
functions of fermions, $1/(e^{E_i/T} + 1)$.

$\nu_R$ production rate $\Gamma_R$ may be compared to
the Hubble rate, 
\begin{eqnarray}
&&
H(T) =\sqrt{\frac{4\pi^3}{45}\, g_*(T)}\, \frac{  T^2}{ m_{\rm pl}}
\,.
\end{eqnarray}
The decoupling temperature $T_d^{\nu_R}$ is defined by equating two rates:
$\Gamma_R(T_d^{\nu_R}) = H (T_d^{\nu_R})$ gives
\begin{eqnarray}
&&
T_d^{\nu_R} =  3.8 \, {\rm MeV}  (\frac{G_R^{-1/2}}{{\rm TeV}})^{4/3} \,
(\frac{g_*^{\nu_R}}{g_*^{\rm EW}})^{1/6}  
\,.
\label {decoupling temp}
\end{eqnarray}
Using the cited value $\sim 550 $MeV of Planck-BAO limit \cite{neff-nur1},
one may derive a limit on $G_R^{-1/2} > 
42 \, $TeV $(g_*^{\nu_R}/g_*^{\rm EW} )^{-1/8}$.

The decoupling temperature of right-handed neutrinos is discussed in the literature \cite{neff-nur1, neff-nur2,neff-nur3, neff-nur4}.
Among them, Ref.~[7] treats of the DLRSM as in the present work.
It is difficult to compare our decoupling temperature with that in Ref.~\cite{neff-nur3},
since it has not given details of how to calculate $\nu_R$ annihilation rate.
We, however, obtain a considerably smaller decoupling temperature than
that of Ref.~\cite{neff-nur3} by a few orders of magnitude.

$\nu_R$ contributes to an extra $\Delta N_{\rm eff}$ given by
\begin{eqnarray}
&&
\frac{\Delta N_{\rm eff}}{3} 
= 
\left(  \frac{ g_*^{\nu_R}}{ g_*^{\nu_L} } \right)^{- 4/3} 
\,,
\label {neff-g* relation 2}
\end{eqnarray}
with $g_*^{\nu_L} = 10.75$ 
(species number at left-handed neutrino decoupling).
If $\nu_R$ decoupling occurs below the electroweak scale 250\,GeV,
one has $\Delta N_{\rm eff} > 0.14$
since 
$ g_*(T^{\rm EW})/g_*^{\nu_L}=9.93 $,
 a value slightly smaller than
Planck limit, but much larger than CMB-S4 target value 0.028.
One can eliminate $g_*^{\nu_R}$ in favor of measurable $\Delta N_{\rm eff}$,
to give
\begin{eqnarray}
&&
T_d^{\nu_R} = 2.95 \, {\rm MeV} (\frac{G_R^{-1/2}}{{\rm TeV}})^{4/3} \,
\Delta N_{\rm eff}^{-1/8}
\,.
\label {td vs dneff}
\end{eqnarray}

Calculations so far are based on that thermal environment consists of
fundamental quarks and leptons.
This picture is valid at temperatures above $\sim 200\,$MeV at which
hadronization occurs and quarks are incorporated into hadrons, 
namely baryons and mesons.
This restricts the applicable region of unification scale to
\begin{eqnarray}
&&
G_R^{-1/2} >  20
\, {\rm TeV}  \, (\frac{  g_*^{\rm EW}}{ g_*^{\nu_R}})^{1/8}
\,.
\end{eqnarray}
For $G_R^{-1/2}$ outside this region the predicted $\Delta N_{\rm eff}$
is grossly inconsistent with Planck observations.
The LR symmetric unification, if the Majorana option
is disfavored, thus appears at energy scale above
$O(10^4)\,$ GeV, far beyond the electroweak scale.

\section
{Further unification for sufficient dilution}

We assume that $\nu_R$ decoupling occurs above the electroweak scale,
hence $\Delta g_* =g_*^{\nu_R} - g_*^{\rm EW} > 0$.
A rational for this assumption is that species dilution below the electroweak scale
is much limited, and forthcoming $\Delta N_{\rm eff}$ observations 
would readily reject the scenario of $\nu_R$ decoupling below electroweak scale,
as seen in Table(\ref{table 1}).

There exists an obvious inequality $\Delta g_* \leq g_*^{\rm tot} - g_*^{\rm EW}$,
with $g_*^{\rm tot}$ the total effective relativistic degrees of freedom
in an extended gauge theory, resulting in $\Delta g_* \leq 18$ and corresponding
$\Delta N_{\rm eff} = 0.114$ in the minimum LR symmetric model.
Thus, if a value of $\Delta N_{\rm eff} < 0.1 $ is measured,
this would exclude the minimum LR symmetric model of Dirac-type
neutrino.

A high energy scale decoupling 
by imposing $T_d^{\nu_R} > 250\,$GeV
(electroweak scale) in (\ref{td vs dneff}) leads to
\begin{eqnarray}
&&
G_R^{-1/2} >  5.0 \times 10^3 \,  {\rm TeV}\, (\Delta N_{\rm eff})^{ 3/32}
\,,
\end{eqnarray}
giving the right-hand side limit, $3.7 \times 10^3
 \,$ TeV for $\Delta N_{\rm eff}=0.05$
(twice of CMB-S4 target value), 
corresponding to
$g_*^{\nu_R} = 230 $ from (\ref{neff-g* relation 2}), a nearly doubled species value
of standard electroweak theory.

\vspace{0.5cm}
Note that the Planck $1\,\sigma$ limit of $\Delta N_{\rm eff} $ is 0.16.
The minimum SU(2)$_R \times$SU(2)$_L \times $U(1) model
gives too small $\Delta g_*$.
It is interesting that measured $\Delta N_{\rm eff}$, hence theoretically inferred 
$\Delta g_*$
from a improved measured value, can determine how large one should think of in
terms of degrees of freedom in Dirac-type neutrino models.

Since  no sizable dilution factor is expected in the DLRSM model,
one needs a proliferated particle spectrum beyond the electroweak scale.
There are a few possibilities: minimum supersymmetric extention of standard model
(MSSM) \cite{mssm 1,mssm}, grand unified theories (GUT) and supersymmetric GUT \cite{susy so(10)}.
For simplicity we assume that there is only one new energy scale beyond
the electroweak scale. The important question is where the new energy scale lies
and how much the dilution factor is.

\begin{figure*}[htb]
\begin{center}
\epsfxsize=0.4\textwidth \epsfbox{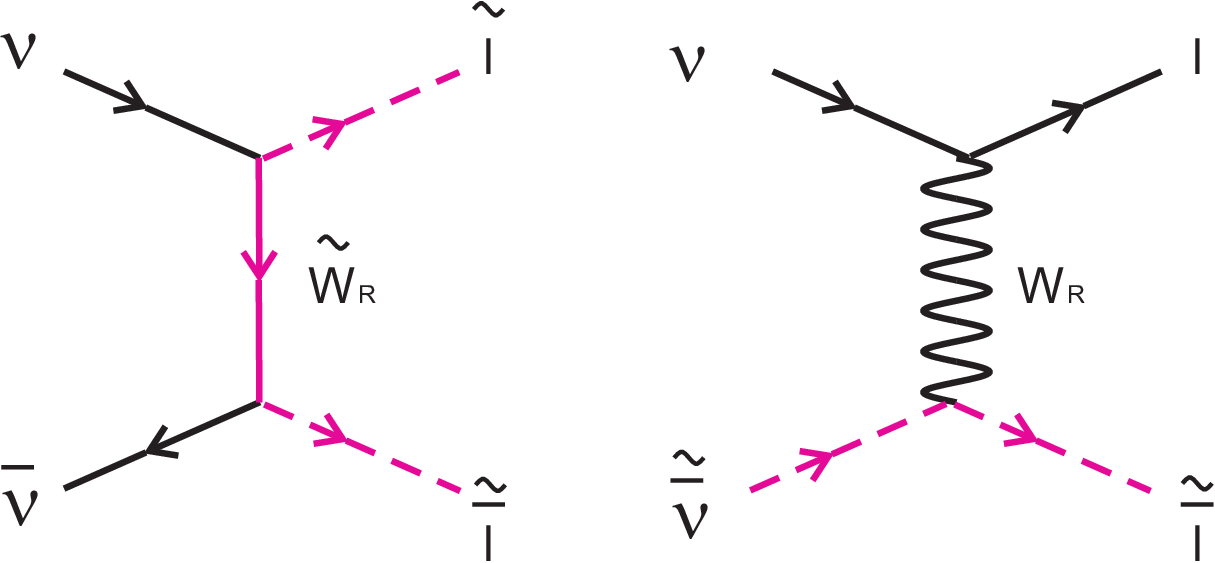}
\vspace{0.5cm}
       \caption{
Examples of SUSY Feynman diagrams.
Super partners are shown in magenda.
             }
       \label{feynman mssm}
\end{center}
\end{figure*}

We focus on the MSSM and assume for simplicity a single
energy scale $M_S$ where all super-partners of standard model particles emerge. 
Total $g_* = 228.75$ in MSSM, $2.143 \times $ standard model value.
In order to estimate $\nu_R$ decoupling temperature, it is necessary to
incorporate new contributions to $\nu_R$ annihilation rate caused by
super-partners.
R-parity conservation in SUSY restricts the number of contributing diagrams:
there are two additional diagrams to each of classified
DLRSM models given in (\ref{diagram 1}) $\sim$
(\ref{diagram 5})  that contain two super-partners
either in final or initial states.
Examples of MSSM diagrams are shown in Fig(\ref{feynman mssm}).

Kinematic variable dependence of cross sections may differ for massless bosons and
massless fermions, for instance in differential cross sections.
Summation over initial thermal particles is also different for bosons and fermions
due to their equilibrium distribution functions.
Incorporating this change of rate in MSSM gives $\nu_R$ decoupling temperature
modified by $O(3^{-1/3 })$, and the new energy scale
\begin{eqnarray}
&&
G_R^{-1/2} > 4.9 \times 10^3\, {\rm TeV} 
(\frac{\Delta N_{\rm eff}}{0.05})^{3/32} (\frac{X}{3})^{1/4}
\,,
\label{eq:def-X}
\end{eqnarray}
with $X$ of order $3$, but expected to deviate slightly.
Thus, the number of diagrams in MSSM are essentially tripled.

\begin{figure*}[htb]
\begin{center}
\epsfxsize=0.6\textwidth \epsfbox{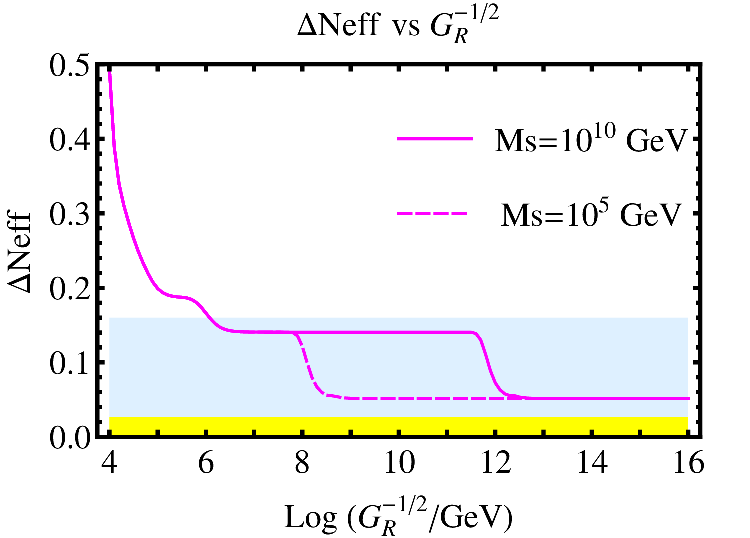}%{Neff-MWR-ver2-fig40.eps}
       \caption{
       Plot of $\Delta N_{\rm eff}$ vs the logarithm of $G_R^{-1/2}/{\rm GeV}$. 
       MSSM particles of a single mass 
 $M_s=10^{10}$ GeV in solid magenda and $10^5$ GeV in
 dotted magenta are taken, and species nearly doubled from the standard model are assumed 
       together with $X=3$ in (\ref{eq:def-X}). 
       The blue band is for Planck 2018 at $1 \sigma$ upper bound, 
       while the yellow band is for $\Delta N_{\rm eff}= 0.028$ CMB-S4 expectation.
       }
       \label{neff vs delg}
\end{center}
\end{figure*}

Suppose that CMB-S4 did not find new contributions and set an
upper limit $\Delta N_{\rm eff} < 0.028$ at 1$\sigma$ level.
As shown in Fig(\ref{neff vs delg}), MSSM extension of DLRSM 
cannot save models of Dirac-type neutrinos.
Nonetheless, Majorana-type neutrino models are acceptable.
Similar arguments in SO(10) models \cite{babu-mohapatra_so(10)}
 based on proliferated species of $g_* = 710$ show that GUT scale of
order $10^{16}$GeV is acceptable to accommodate DLRSM structure.
Thus, CMB-S4 is expected to have a great impact on physics beyond standard model.

\section
{Combined analysis using summed neutrino mass measurement}

Extra $\Delta N_{\rm eff}$ due to relic $\nu_R$ contributes to
the summed neutrino masses, which is also a target
of forthcoming CMB-S4 observations.

\begin{figure*}[htb]
\begin{center}
       \includegraphics[clip,bb=0 0 1000 800, width=8.0cm]{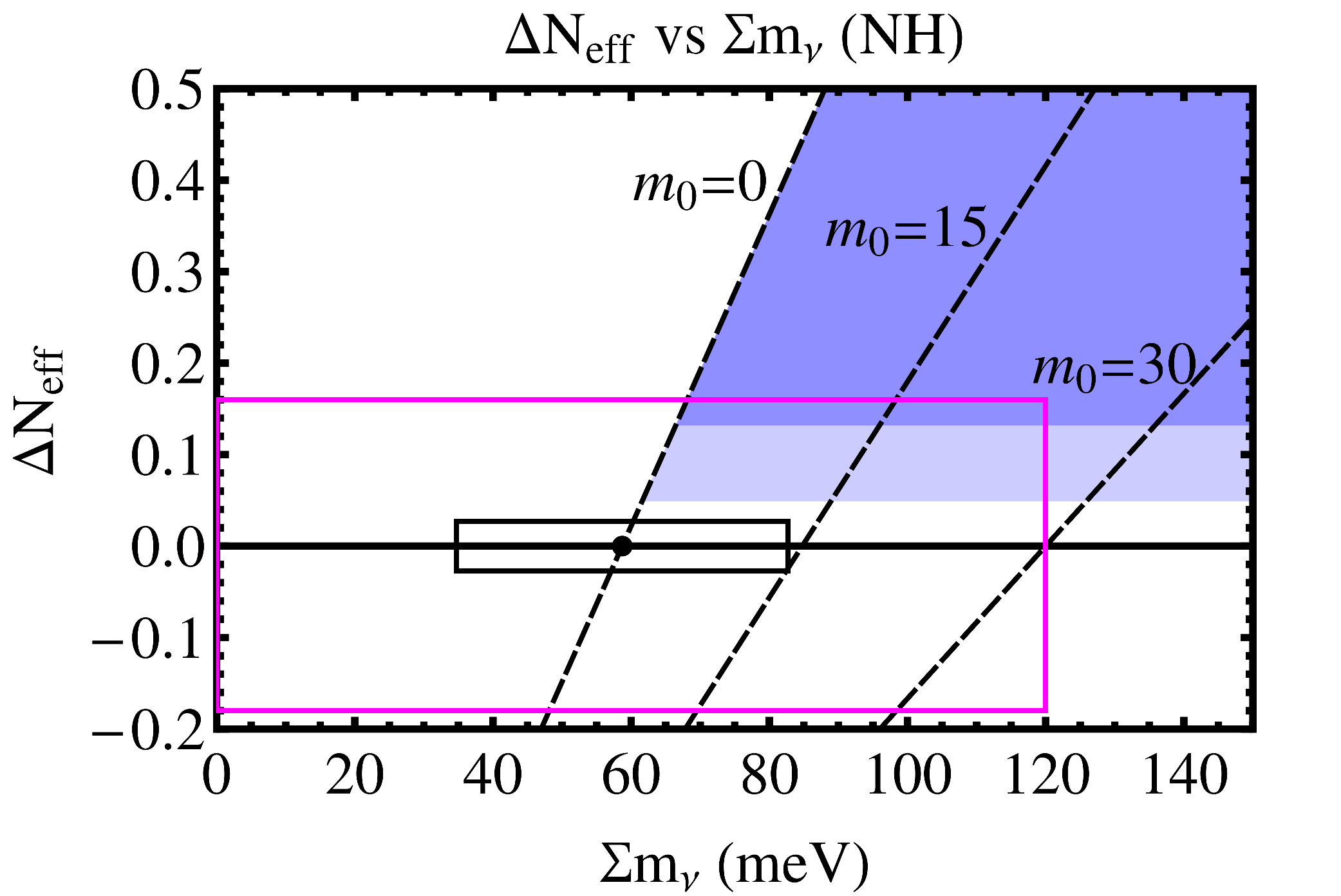}
       \includegraphics[clip,bb=0 0 1000 800, width=8.0cm]{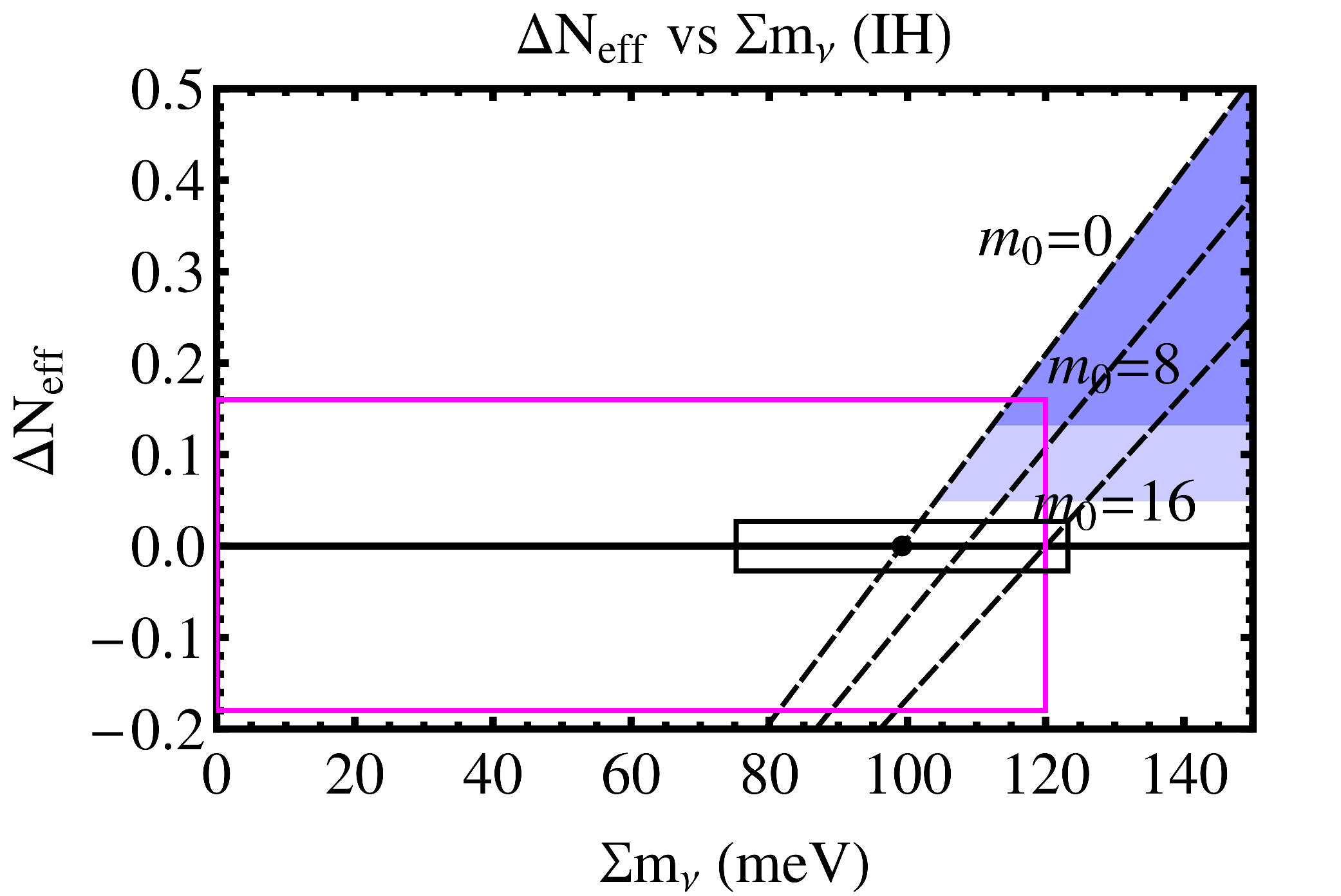}
\caption{
      Plot of $\Delta N_{\rm eff}$ vs $\sum m_{\nu}$   
       for NH (left panel) and IH (right panel). 
       The large rectangular box in  magenta indicates  
       the current $1 \sigma$ limits on $\Delta N_{\rm eff}$ and $\sum m_{\nu}$ 
       determined by Planck 2018 \cite{planck-2018}. 
       The small rectangular box in black shows the anticipated $\pm 1 \sigma$ errors by future observations \cite{cmb-s4}. 
       The center point would move inside or around the large rectangular box depending on the outcome of the observations.
       The three dashed lines are the $\Delta N_{\rm eff}$-vs-$\sum m_{\nu}$ relations, given in Eqs.~\eqref{Eq:NH} and \eqref{Eq:IH},  
       with $m_0$ indicated in the figure.  
       The dark and light blue shaded areas indicate the allowed regions of
       the DLRSM and its MSSM extension with sufficiently high decoupling temperature,
       respectively.}
       \label {Neff_mass-sum}
\end{center}
\end{figure*} 

It is  of great value to 
summarize analyzed data using the minimum neutrino mass value
defined as $m_0$. 
Neutrino oscillation data  gives
the summed neutrino mass in terms of different functions
of $m_0$,  with different offset parameters   in  the normal hierarchical (NH) 
and the inverted hierarchical (IH) orderings; 
\begin{equation}
\hspace*{-0.2cm}
\frac{\sum_i m_i }{ 1+ \frac{\Delta N_\text{eff}}{3} } =
m_0 + \sqrt{ 8.61^2 + m_0^2}
+ \sqrt{ 50.1^2 + m_0^2} 
\,,\label{Eq:NH}
\end{equation}
for NH, and
\begin{equation}
\hspace*{-0.2cm}
\frac{
\sum_i m_i }{ 1+ \frac{\Delta N_\text{eff}}{3} } =
m_0 + \sqrt{ 50.0^2  + m_0^2}
+ \sqrt{ 49.2^2 + m_0^2} 
\,,\label{Eq:IH}
\end{equation}
for IH, where mass values are given in meV unit.
We note that $\sum_i m_i$ denotes the summed neutrino mass determined by cosmological observations. 
This quantity should be divided by the dilution factor to compare with the values of the oscillation experiments\cite{pdg}.
Offset values given by setting $m_0=0$ are  $> 58.7$ and $> 99.2$ meV, for NH and IH, respectively.
This difference in two ordering schemes is significantly large. 

In Fig(\ref{Neff_mass-sum}) we show the  region 
in the $(\Delta N_{\rm eff}\,, \sum_i m_i)$ plane 
allowed by Planck 2018 observations.
%Results are applicable irrespective of Majorana or Dirac
%type of neutrinos.
These results hold irrespective of whether neutrinos are of Majorana or of Dirac type, since neutrino oscillation data cannot distinguish these two types.
Nevertheless, plots given here can help
NH and IH differences in future cosmological observations.

In the case of Dirac-type neutrino one can add more to this.
One can readily convert $\Delta N_{\rm eff}$ to $ \Delta g_*$
using (\ref{neff-g* relation}) in favor of theoretical convenience
of Dirac-type neutrino.
We illustrate results in Fig(\ref{Neff_mass-sum}) 
for three choices of the smallest neutrino mass.
With improved accuracy in future observations one can reject
IH schemes of larger smallest mass more readily than
NH schemes.
The minimum LR symmetric model is on the verge of
rejection if IH scheme is adopted.

Considering these forecasts of cosmological observations, direct
experiments of the smallest mass measurement and M/D  distinction
in terrestrial laboratories becomes even more important.
 It has been proposed to
use laser-initiated coherence to measure these in atomic experiments
in order to enhance otherwise tiny rates
\cite{fukumi}.

\vspace{1cm}
In summary, we investigated 
how cosmological observations of neutrino properties can probe
still undetermined neutrino mass types
and measure mass parameters with precision, under a few plausible assumptions:
(1) three neutrino scheme, (2) zero chemical potential of thermal particles,
(3) no hypothetical light relic 
or relics with small
 accumulated contribution to $\Delta N_{\rm eff}$ less than $\sim 0.05$.

The Majorana neutrino, with rapid decay of its heavy right-handed partners,
 is in  agreement with cosmological observations at nucleo-synthesis
and at later epochs after recombination.
But, Dirac-type right-handed neutrinos $\nu_R$'s,
must be diluted away, to give
$\Delta N_{\rm eff} < 0.16 $, as limited by Planck + DESI observations.
The necessary dilution is provided by rehearing left-handed neutrino
after cosmological $\nu_R$ decoupling, hence the problem is sensitive
to particle content in thermal equilibrium.
It is necessary, for the answer to this question, to 
identify a proper theoretical framework of
how right-handed neutrino bound is satisfied.
We find it most natural to adopt
gauge theories including  SU(2)$_R \times $SU(2)$_L \times $U(1)
as a subgroup, and how a nearly complete dilution of $\nu_R$
may occur, requiring a prolific set of particles as large as $O(100)$.

Implication of CMB-S4 sensitivity $\Delta N_{\rm eff} \sim 0.028$
in Dirac-type neutrino models
is that 
either SU(2)$_R \times $SU(2)$_L \times $U(1) is extended by
supersymmetry allowing $\sim 2$ dilution after decoupling or
grand unified extension of the left-right symmetric
models is required as typical favored cases, thus making it possible to explore
highest energy scale of particle physics.

The choice of NH or IH mass ordering scheme, irrespective of the mass types,
 should not be too difficult to determine in CMB-S4 observations.

\vspace{1cm}
Note added.

After completion of this work we became aware of the work,
DESI Collaboration, 
arXiv:2404/03002v2[astro-ph[CO],
in which analysis of DESI observations suggests that
72 meV is likely to be the upper bound of summed neutrino mass.
This makes the right panel  IH case in our Fig(\ref{Neff_mass-sum})
disfavored.

\begin{acknowledgments}
We appreciate O. Tajima at Kyoto University for valuable information and comments on CMB-S4 and other projects.
This research was partially supported by Grant-in-Aid  
19H00686 (NS), 18K03621 (MT), and 21K03575 (MY) from the Japanese Ministry of Education, Culture, Sports, Science, and Technology.
\end{acknowledgments}

\appendix

\section{Doublet left-right symmetric model}

\subsection{Gauge symmetry and matter contents}
The gauge symmetry of doublet left-right symmetric model (DLRSM) is 
\begin{eqnarray}
\text{SU}(3)_\text{C}\times\text{SU}(2)_\text{L}\times\text{SU}(2)_\text{R}
\times\text{U}(1)_\text{B-L}
\,. 
\end{eqnarray}

The fermions in DLRSM are
\begin{eqnarray}
&&Q_L=\begin{pmatrix}u_L\\ d_L\end{pmatrix}\,,\ 
 Q_R=\begin{pmatrix}u_R\\ d_R\end{pmatrix}\,,\\
&&L_L=\begin{pmatrix}\nu_L\\ \ell_L\end{pmatrix}\,,\ 
 L_R=\begin{pmatrix}\nu_R\\ \ell_R\end{pmatrix}\,,
\end{eqnarray}
where gauge quantum numbers of underlying group are $(3,2,1,1/3)$,
$(3,1,2,1/3)$, $(1,2,1,-1)$, $(1,1,2,-1)$,
respectively for quarks $Q_L$, $Q_R$ and leptons $L_L$, $L_R$.

Three types of scalars in DLRSM are introduced:
\begin{eqnarray}
&&
\Phi=\begin{pmatrix}
       \phi_1^0 & \phi_2^+\\
       \phi_1^- & \phi_2^0
      \end{pmatrix}:\ (1,2,2,0)\,,
\\ &&
\chi_L=\begin{pmatrix}\chi_L^+\\ \chi_L^0\end{pmatrix}:\ (1,2,1,1)\,,
\\ &&
 \chi_R=\begin{pmatrix}\chi_R^+\\ \chi_R^0\end{pmatrix}:\ (1,1,2,1)\,.
\end{eqnarray}

\subsection{Spontaneous symmetry breaking (SSB)}
The vacuum expectation values
\begin{eqnarray}
&&
\langle\Phi\rangle=\begin{pmatrix}
         v_1/\sqrt{2} & 0\\
         0 & v_2/\sqrt{2}
        \end{pmatrix}\,,\ 
\\ &&
\langle\chi_L\rangle=\begin{pmatrix}0 \\ v_L/\sqrt{2}\end{pmatrix}\,,\ 
\\ &&
\langle\chi_R\rangle=\begin{pmatrix}0 \\ v_R/\sqrt{2}\end{pmatrix}\,,
\end{eqnarray}
with a fine-tuning inequality $v_R\gg v_1,v_2,v_L$ lead to the following SSB pattern:
\begin{eqnarray}
&&
\text{SU}(2)_\text{L}\times\text{SU}(2)_\text{R}\times\text{U}(1)_\text{B-L}
\stackrel{v_R}\longrightarrow
\text{SU}(2)_\text{L}\times\text{U}(1)_\text{Y}
\nonumber \\ &&
\stackrel{v_{1,2,L}}
\longrightarrow
\text{U}(1)_\text{em}\,.
%\nonumber \\ &&
\end{eqnarray}
The first step is relevant to the study of $\nu_R$ decoupling above
the electroweak scale $v^2=v_1^2+v_2^2+v_L^2\simeq (246\ \text{GeV})^2$.

With a fine-tuning $v \ll v_R$  there is only one light Higgs boson $h$ of mass in the
electroweak energy scale, while all other Higgs bosons remain
much heavier than electroweak scale.

\subsection{Nonstandard gauge bosons}
In addition to the gauge bosons in the standard model,
we have a pair of charged gauge bosons $W^\pm_{R\mu}$ and a neutral gauge
boson $Z_{R\mu}$ in the DLRSM. 
The mass eigenstates are expressed in terms of the gauge basis field as
\begin{eqnarray}
&& W^\pm_{R\mu}=\frac{1}{2}(W_{R\mu}^1\mp W_{R\mu}^2)\,,\\
&& Z_{R\mu}=\frac{1}{g_{Z_R}}(g_R W_{R\mu}^3-g_\text{B-L} B_{\text{B-L}\mu})\,,\\
&& g_{Z_R}=\sqrt{g_R^2+g_\text{B-L}^2}\,,
\end{eqnarray}
where $g_R$ and $g_\text{B-L}$ are the gauge coupling constants of 
$\text{SU}(2)_\text{R}$ and $\text{U}(1)_\text{B-L}$ respectively.
Their masses are given by
\begin{eqnarray}
M_W=\frac{1}{2}g_R v_R\,,\
M_Z=\frac{1}{2}g_{Z_R} v_R\,.
\end{eqnarray}

Incidentally, the $\text{U}(1)_\text{Y}$ gauge boson is identified
as the orthogonal state of $Z_R$ as
\begin{eqnarray} 
B_\mu=\frac{1}{g_{Z_R}}(g_\text{B-L} W_{R\mu}^3+g_R B_{\text{B-L}\mu})\,,
\end{eqnarray}
and it is massless at this stage.
With the gauge mixing angle defined by
\begin{eqnarray}
\tan\theta_R=\frac{g_\text{B-L}}{g_R}\,,
\end{eqnarray}
one may express the neutral gauge bosons in the gauge basis 
in terms of those in the mass basis as
\begin{eqnarray}
&& W_{R\mu}^3=Z_{R\mu}\cos\theta_R+B_\mu\sin\theta_R\,,\\
&& B_{\text{B-L}\mu}=-Z_{R\mu}\sin\theta_R+B_\mu\cos\theta_R\,.
\end{eqnarray}

\subsection{Nonstandard gauge interactions of fermions}
The kinetic term of fermions is
\begin{eqnarray}
\mathcal{L}_f=\bar f i\slashed{D}f\,,
\end{eqnarray}
with the covariant derivative $D_{\mu}$ given by
\begin{eqnarray}
D_\mu =&&\partial_\mu+i g_L T_L^a W_{L\mu}^a+i g_R T_R^a W_{R\mu}^a\nonumber\\
       &&+i \frac{g_\text{B-L}}{2}(B-L) B_{\text{B-L}\mu}\,,\nonumber\\
      =&&\partial_\mu+i g_L T_L^a W_{L\mu}^a + i g_Y Y B_\mu\nonumber\\
       && +i\frac{g_R}{\sqrt{2}}(T_R^+ W_{R\mu}^+ + T_R^- W_{R\mu}^-)\nonumber\\
       &&+i g_{Z_R}(T_R^3-Y\sin^2\theta_R)Z_{R\mu}\,,
\end{eqnarray}
where $T_R^\pm=T_R^1\pm i T_R^2$. The $W_R$ interaction is the one in
the standard model with the left-handed fermions replaced by 
the right-handed ones. 
We find the $Z_R$ interaction as
\begin{eqnarray}
\mathcal{L}_{Z_R\bar f f}=-g_{Z_R}Z_{R\mu}(c_R^f\bar f_R\gamma^\mu f_R+
                                       c_L^f\bar f_L\gamma^\mu f_L)\,,
\end{eqnarray}
where
\begin{eqnarray}
c_R^f=T_R^3-Y\sin^2\theta_R\,,\ c_L^f=-Y\sin^2\theta_R\,,
\end{eqnarray}
where $Y(\nu_R)=0$, $Y(\ell_R)=-1$, $Y(L_L)=-1/2$, $Y(u_R)=2/3$,
$Y(d_R)=-1/3$ and $Y(Q_L)=1/6$.

\subsection{Gauge coupling constants}
The nonstandard gauge coupling constants $g_R$ and $g_\text{B-L}$ are related
to $g_Y$ by
\begin{eqnarray}
\frac{1}{g_R^2}+\frac{1}{g_\text{B-L}^2}
=\frac{1}{g_Y^2}=\frac{\cos^2\theta_W}{4\pi\alpha}\,.
\end{eqnarray}
Thus, we cannot choose $g_R$ and $\sin\theta_R$ independently. 
Assuming the manifest left-right symmetry $g_R=g_L(=e/\sin\theta_W)$,
and taking $\alpha=1/128$ and $\sin^2\theta_W=0.23$, we obtain
$g_R=0.653$ and $\sin^2\theta_R=0.299$.

\section{Right-handed neutrino annihilation processes}

Processes are listed in the text.
We shall give amplitudes and squared amplitudes when relevant
helicity states are added.

\subsection{Amplitudes}
\subsubsection{t-channel $W_R$ exchange $(a)$}
\begin{eqnarray}
&&
\mathcal{M}_{t1}
=\frac{g_R^2}{2}\frac{1}{t-M_W^2}\mathcal{A}^{(1)}\,,
\\ &&
 \mathcal{A}^{(1)}=
  \bar u(q_1)\gamma^\alpha P_R u(p_1)\bar v(p_2)\gamma_\alpha P_R v(q_2)\,,
\end{eqnarray}
where $t=(p_1-q_1)^2=(q_2-p_2)^2$.

\subsubsection{t-channel $W_R$ exchange $(a)'$}
\begin{eqnarray}
&&
\mathcal{M}_{t2}
=\frac{g_R^2}{2}\frac{1}{t-M_W^2}\mathcal{A}^{(2)}\,,
\\ &&
 \mathcal{A}^{(2)}=
  \bar u(q_1)\gamma^\alpha P_R u(p_1)\bar u(q_2)\gamma_\alpha P_R u(p_2)\,.
\end{eqnarray}

\subsubsection{s-channel $Z_R$ exchange $(b)$, $(b)'$}
\begin{eqnarray}
&&
\mathcal{M}_{s,f_H}
=g_{Z_R}^2 c_R^\nu c_H^f\frac{1}{s-M_Z^2}\mathcal{A}^{(3)}_H\,,
\\ &&
 \mathcal{A}^{(3)}_H=
   \bar v(p_2)\gamma^\alpha P_R u(p_1)\bar u(q_1)\gamma_\alpha P_H v(q_2)\,,
\end{eqnarray}
where 
$s=(p_1+p_2)^2=(q_1+q_2)^2$, $P_{R/L}=(1\pm\gamma_5)/2$ and $H=R,L$.

\subsubsection{s-channel $W_R$ exchange $(c)$}
\begin{eqnarray}
\mathcal{M}_{s}
=\frac{g_R^2}{2}\frac{1}{s-M_W^2}\mathcal{A}^{(3)}_R\,.
\end{eqnarray}

\subsection{Squared amplitudes}
We evaluate $\overline{\sum}|\mathcal{M}|^2$, where $\overline{\sum}$ means
the average of initial spins and sum over final spins in spinor manipulation.
It is straight forward to obtain the following squared fundamental amplitudes:
\begin{eqnarray}
&&\overline{\sum}|\mathcal{A}^{(1)}|^2=(s+t)^2\,,\\
&&\overline{\sum}|\mathcal{A}^{(2)}|^2=s^2\,,\\
&&\overline{\sum}|\mathcal{A}^{(3)}_R|^2=(s+t)^2\,,\\
&&\overline{\sum}|\mathcal{A}^{(3)}_L|^2=t^2\,.
\end{eqnarray}
We also need the following interference term:
\begin{eqnarray}
&\overline{\sum}\mathcal{A}^{(3)}_R\mathcal{A}^{(1)*}
 =\overline{\sum}\mathcal{A}^{(3)*}_R\mathcal{A}^{(1)}=-(s+t)^2\,.
\end{eqnarray}

These results of helicity-summed squared
amplitudes are checked by the software FeynCalc\cite{FeynCalc} as well.

\subsection{Cross sections}
The differential cross section is expressed by
\begin{eqnarray}
d\sigma=\frac{1}{2s\bar\beta_i}\overline{\sum}|\mathcal{M}|^2 d\Phi_2\,,
\end{eqnarray}
where the two-body phase space is given by
\begin{eqnarray}
d\Phi_2=&&(2\pi)^4\delta^4(p_1+p_2-q_1-q_2)\nonumber\\
        &&\frac{d^3q_1}{2q_1^0(2\pi)^3}\frac{d^3q_2}{2q_2^0(2\pi)^3}\,,\\
\int d\Phi_2=&&\frac{dt}{8\pi s\bar\beta_i}\,.
\end{eqnarray}
We have performed trivial parts of integration leaving the integration over $t$ variable. 
In the case of massless particles, $\bar\beta_i=1$ 
and the range of integration is $-s\leq t\leq 0$.

\subsubsection{Right-handed neutrino pair annihilation process $(al)$, $(b)$}
%$\nu^\ell_R\bar\nu^\ell_R\to \ell_R\bar\ell_R$:
\begin{eqnarray}
&&
\sigma^{(1)}=
 \frac{1}{16\pi s}
\left[\frac{1}{3}g_{Z_R}^4c_R^{\nu 2}c_R^{\ell 2}\left(\frac{s}{s-M_Z^2}\right)^2\right.
\nonumber\\ 
&&+\frac{g_R^4}{4}\left\{2+\frac{s}{M_W^2}-2\left(1+\frac{M_W^2}{s}\right)
                         \log\left(1+\frac{s}{M_W^2}\right)
                  \right\}\nonumber\\
&&-g_R^2g_{Z_R}^2c_R^{\nu}c_R^{\ell}\frac{s}{s-M_Z^2}
   \left\{\frac{3}{2}+\frac{M_W^2}{s}\right.\nonumber\\
&&\left.\left.
  -\left(1+\frac{M_W^2}{s}\right)^2\log\left(1+\frac{s}{M_W^2}\right)\right\}
\right]\,.
\end{eqnarray}
In the four-Fermi approximation $M_W\gg s$, using 
$M_W=M_Z\cos\theta_R$ and $g_R=g_{Z_R}\cos\theta_R$ (namely $\rho_R=1$),
we find
\begin{eqnarray}
\sigma^{(1)}_\text{4F}=
 \frac{2}{3\pi}G_R^ 2s\left(c_R^\nu c_R^\ell+\frac{1}{2}\right)^2,\ 
\frac{G_R}{\sqrt{2}}=\frac{g_R^2}{8 M_W^2}\,.
\end{eqnarray}

\subsubsection{Right-handed neutrino pair annihilation process $(b)$, $(b)'$}
\begin{eqnarray}
\sigma^{(2)}=&&\frac{n_c}{48\pi s}g_{Z_R}^4 c_R^{\nu 2}c_H^{f2}
                \left(\frac{s}{s-M_Z^2}\right)^2\,,\nonumber\\
             &&H=L,R\,,\\
\sigma^{(2)}_\text{4F}=&&\frac{2 n_c}{3\pi}G_R^2 s c_R^{\nu 2}c_H^{f2}\,,
\end{eqnarray}
where $n_c(=3)$ represents the number of colors.

\subsubsection{Single right-handed neutrino annihilation process
$(c)$}
\begin{eqnarray}
&&\sigma^{(3)}=\frac{n_c}{48\pi s}\frac{g_R^4}{4}
       \left(\frac{s}{s-M_W^2}\right)^2\,,\\
&&\sigma^{(3)}_\text{4F}=\frac{n_c}{6\pi}G_R^2 s\,.
\end{eqnarray}

\subsubsection{Single right-handed neutrino annihilation process
$(al)$, $(aq)$}
\begin{eqnarray}
\sigma^{(4)}=&&\frac{1}{16\pi s}\frac{g_R^4}{4}
                \left[2+\frac{s}{M_W^2}\right.\nonumber\\
             &&\left.-2\left(1+\frac{M_W^2}{s}\right)
                \log\left(1+\frac{s}{M_W^2}\right)\right]\,,\\ 
\sigma^{(4)}_\text{4F}=&&\frac{1}{6\pi}G_R^2 s\,.
\end{eqnarray}

\subsubsection{Single right-handed neutrino annihilation process
$(a)'$}
\begin{eqnarray}
&&\sigma^{(5)}=\frac{1}{16\pi s}\frac{g_R^4}{4}\frac{s^2}{M_W^4+s M_W^2}\,,\\
&&\sigma^{(5)}_\text{4F}=\frac{1}{2\pi}G_R^2 s\,.
\end{eqnarray}

\subsubsection{Total cross section}
To summarize, the total $\nu^\ell_R$ annihilation cross section is given by
\begin{eqnarray}
\sigma_\text{4F}=&&\frac{G_R^2s}{6\pi}\left[5 n_c n_g+n_g +4c_R^\nu\left\{-c_R^\ell+n_g
c_R^\nu\right.\right.\nonumber\\
 &&\left.\left.\left(c_L^{\ell 2}+c_L^{\nu 2}+c_R^{\ell 2}
                     +n_c(c_L^{d2}+c_L^{u2}+c_R^{d2}+c_R^{u2})\right)
    \right\}\right]\,,\nonumber\\
=&&\frac{2G_R^2s}{\pi}I(\sin^2\theta_R)\,,
\end{eqnarray}
where $I(x):=(217-56x+40x^2)/48$, and
$n_g(=3)$ represents the number of generations.

\section{Thermal average}
We consider the thermal average of cross section $\sigma$ times 
(M{\o}ller) velocity $v$:
\begin{eqnarray}
&&
\langle\sigma v\rangle=\frac{g_1g_2}{n_1n_2}
 \int\sigma v f(\bm{p}_1)f(\bm{p}_2)
     \frac{d^3p_1}{(2\pi)^3}\frac{d^3p_2}{(2\pi)^3}\,,
\\ &&
 v:=\frac{\sqrt{(p_1\cdot p_2)^2-m_1^2m_2^2}}{E_1E_2}\,,
\end{eqnarray}
where $g_{1,2}$, $n_{1,2}$ and $f(\bm{p}_{1,2})$ denote the spin degrees of 
freedom, the number densities and the thermal distributions of the initial 
particles respectively. For the case of massless initial particles,
we find $v=s/(2E_1E_2)$ and
\begin{eqnarray}
n&&=g\int f(\bm{p})\frac{d^3p}{(2\pi)^3}\nonumber\\
 &&=\frac{3}{4}\zeta(3)\frac{g}{\pi^2}T^3\simeq 0.90\frac{g}{\pi^2}T^3\,,\ \text{Fermi-Dirac}\,,\\
 &&=\frac{g}{\pi^2}T^3\,,\  
 \text{Maxwell-Boltzmann}\,.
\end{eqnarray}

Relation of Fermi-Dirac (FD) and approximate
Maxwell-Boltzmann (MB) distribution functions is
$f^{\rm FD}= 1/(e^{E/T} + 1) \rightarrow f^{\rm MB} = e^{-E/T} $.

\subsection{Maxwell-Boltzmann approximation \\
for initial phase space integration}
When the cross section is suppressed for smaller $s$ as in the case of
the four-Fermi interaction, we expect that the Maxwell-Boltzmann distribution
is a good approximation to the Fermi-Dirac distribution in the thermal average.

To evaluate the thermal integration, the following change of variables 
is convenient\cite{GondoloGermini1991}:
\begin{eqnarray}
d^3p_1d^3p_2=2\pi^2E_1E_2dE_+dE_-ds
\,,
\end{eqnarray}
where $E_\pm=E_1\pm E_2$ and the integration region is
\begin{eqnarray} 
s>0\,,\ E_+>\sqrt{s}\,,\ |E_-|<\sqrt{E_+^2-s}\,,
\end{eqnarray}
for the massless case. With the Maxwell-Boltzmann distribution
$f(\bm{p})=e^{-E/T}$, we obtain (for the massless case)
\begin{eqnarray}
\langle\sigma v\rangle_\text{MB}=\frac{g_1g_2}{n_1n_2}
 \frac{T^2}{32\pi^4}\int\sigma s\frac{\sqrt{s}}{T}K_1(\sqrt{s}/T)ds\,,
\end{eqnarray}
where $K_n(z)$ represents the modified Bessel function of the second kind.

\subsubsection{Four-Fermi approximation}
As explicitly shown in the previous section, the cross section
is proportional to $s$ in the four-Fermi approximation.
We find that  the relevant thermal average is given by
\begin{eqnarray}
\langle s v\rangle_\text{MB}=\frac{g_1g_2}{n_1n_2}\frac{24}{\pi^4}T^8
                  =24T^2\,,
\end{eqnarray}
and
\begin{eqnarray}
\hspace*{-0.3cm}
\langle\sigma_\text{4F} v\rangle_\text{MB}=
 \frac{48G_R^2T^2}{\pi}I(\sin^2\theta_R)\,.
\end{eqnarray}

\subsection{Thermal average with Fermi-Dirac distribution}
%\subsubsection{Four-Fermi approximation}
The relevant thermal average using Fermi-Dirac distribution function is
\begin{eqnarray}
\langle s v\rangle_\text{FD}
 =\frac{g_1g_2}{n_1n_2}\frac{T^8}{32\pi^4}I_\text{FD}
\end{eqnarray}
where 
\begin{eqnarray}
I_\text{FD}=&&\int d\frac{s}{T^2}d\frac{E_1}{T}\left(\frac{s}{T^2}\right)^2\nonumber\\
            &&\ \ \frac{\log\left(1+e^{-s/4(E_1/T)T^2}\right)}{e^{E_1/T}+1}\,,\\
           =&&\int_0^\infty dy y^2\int_0^\infty dx
              \frac{\log(1+e^{-y/4x})}{e^x+1}\,.
\end{eqnarray}
Exchanging the order of $x$ and $y$ integrals, we obtain analytic result
given in the text,
\begin{align}
I_\text{FD}=\frac{49\pi^8}{675}\simeq 688.80\,.
\end{align}
Then, we find
\begin{eqnarray}
&&
\langle\sigma_\text{4F} v\rangle_\text{FD}=
 \left(\frac{4}{3\zeta(3)}\right)^2\frac{I_\text{FD}}{768}
 \langle\sigma_\text{4F} v\rangle_\text{MB}
\nonumber \\ &&
 =\frac{49\pi^8}{291600\zeta(3)^2}\langle\sigma_\text{4F}v\rangle_\text{MB}
 \simeq 1.1035 \langle\sigma_\text{4F} v\rangle_\text{MB}\,.
\end{eqnarray}

\section{Decoupling temperature}
The right-handed neutrino annihilation rate is given by
$\Gamma=4n_{\nu^\ell_R}\langle\sigma v\rangle$.
We note that the factor of $4=2^2$ is introduced to compensate
the initial spin average factor in the cross sections given in this appendix.
In the four-Fermi and Maxwell-Boltzmann approximation, we obtain
\begin{eqnarray}
\hspace*{-0.3cm}
\Gamma_\text{4F,MB}=\frac{192}{\pi^3}G_R^2T^5I(\sin^2\theta_R)\,.
\end{eqnarray}
With the Fermi-Dirac distribution, we find
\begin{eqnarray}
&&
\Gamma_\text{4F,FD}
=\frac{4}{3\zeta(3)}\frac{I_\text{FD}}{768}\Gamma_\text{4F,MB}
=\frac{49\pi^8}{388800\zeta(3)}\Gamma_\text{4F,MB}
\nonumber \\ &&
\simeq 0.99482\Gamma_\text{4F,MB}\,.
\end{eqnarray}
It turns out that the the Maxwell-Boltzmann approximation is surprisingly
accurate.
To summarize,
\begin{align}
\Gamma_\text{4F,FD(MB)}=
 C_\text{4F,FD(MB)}G_R^2T^5I(\sin^2\theta_R)\,,
\end{align}                       
where
\begin{align}
C_\text{4F,FD}=\frac{49 \pi^5}{2025\zeta(3)}\,,\ 
C_\text{4F,MB}=\frac{192}{\pi^3}\,.
\end{align}
The decoupling temperature is defined by
\begin{eqnarray}
\Gamma(T_\text{dec})=H(T_\text{dec})\,,\ 
H(T)=\sqrt{\frac{8\pi}{3}\frac{\pi^2}{30}g_*}\,\frac{T^2}{m_\text{Pl}}\,,
\end{eqnarray}
where $H(T)$ is the Hubble constant and the Planck mass is 
$m_\text{Pl}=1.2211\times10^{19}\ \text{GeV}$.
We note that $g_*=g_*^\text{SM}=427/4$ for the standard model and
$g_*=g_*^\text{SM}+2\times 3\times (7/8)$ in the present case.

In the four-Fermi approximation, the right-handed neutrino decoupling 
temperature is expressed by
\begin{eqnarray}
\!\!\!\!\!T^{\nu_R}_\text{dec}\!=\!
 \left[\sqrt{\frac{8\pi^3}{90}g_*}
       \frac{1}{C_\text{4F,FD(MD)}G_R^2m_\text{Pl}I(\sin^2\theta_R)}
 \right]^{1/3}\!\!\!\!.
\end{eqnarray}

\end{document}